\documentclass[pra,amsmath,a4paper,twocolumn]{revtex4}
\usepackage{graphicx}
\usepackage{amsfonts}
\usepackage{amsmath}
\usepackage{ltxtable}
\usepackage{CJK}
\usepackage{mdwmath}
\usepackage{color}

\newcommand{\mA}{{\mathcal A}}
\newcommand{\mN}{{\mathcal N}}

\newcommand{\mL}{{\mathcal L}}
\newcommand{\mB}{{\mathcal B}}
\newcommand{\mC}{{\mathbb C}}
\newcommand{\mmC}{{\mathcal C}}
\newcommand{\mZ}{{\mathbb Z}}
\newcommand{\mV}{{\mathbb V}}
\newcommand{\mE}{{\mathbb E}}

\newcommand{\mF}{{\mathbb F}}
\newcommand{\mCP}{{\mathcal {CP}}}
\newcommand{\mFL}{{\mathcal {FL}}}
\newcommand{\mFC}{{\mathcal {FC}}}
\newcommand{\mW}{{\mathcal W}}
\newcommand{\mSL}{{\mathcal {SL}}}
\newcommand{\mTL}{{\mathcal {TL}}}
\begin{document}

\title{Graphical Non-contextual Inequalities for Qutrit Systems}

\author{Weidong Tang$^{1}$
and Sixia Yu$^{2}$}
\affiliation{$^1$Key Laboratory of Quantum Information and Quantum Optoelectronic Devices, Shaanxi Province,
 and Department of Applied Physics
of Xi'an Jiaotong University, Xi'an 710049, P.R. China
\\
$^2$Hefei National Laboratory for Physical Sciences at
Microscale and Department of Modern Physics
of University of Science and Technology of China, Hefei 230026, P.R. China}

\begin{abstract}
One of the interesting topics in quantum contextuality is the construction for various non-contextual inequalities. By introducing a new structure called hyper-graph, we present a general method, which seems to be analytic and extensible, to derive the non-contextual inequalities for the qutrit systems. Based on this,  several typical families of non-contextual inequalities are discussed. And our approach may also help us to simplify some state-independent proofs for quantum contextuality in one of our recent works.

\end{abstract}

\maketitle

\section{Introduction}

It is well known that a violation of a Bell inequality\cite{Bell} can be used for refuting the local realism assumption of  quantum mechanics. More generally, any quantum violation for a non-contextual inequality\cite{KCBS,cabello0801,cabello08,yu-oh,cabello1201,TYO,NS condi} can be used to disprove the non-contextual assumption for  quantum mechanics, and can be considered as another version of the proof for quantum contextuality or the Kochen-Specker(KS) theorem\cite{Bell2,KS,mermin1}.
One of the proofs for the KS theorem is to find a contradiction of a KS value assignment --- which claims that the value assignment to an observable (can only be assigned to one of its eigenvalues) is independent of the context it measured alongside --- to a set of chosen rays. And a non-contextual inequality can be considered as a bridge to connect a logical proof of the KS theorem and a corresponding experimental verification\cite{Lapkiewicz,Zu,Vincenzo,XiangZhang,Huang2}.

Recently, to give a universal construction for the state-independent proof for quantum contextuality, we introduced a $(6n+2)$-ray model\cite{Tang-yu}.  As a kind of special state-dependent proofs for quantum contextuality, this family of models, can induce a type of basic non-contextual inequalities. Based on this, we can analytically derive numerous non-contextual inequalities.

In this paper, starting with the $(6n+2)$-ray model,  we  introduce a kind of generalized graph which is called a `` hyper-graph". Then we give several interesting families of non-contextual inequalities from each kind of hyper-graphs. Finally, we give a rough analysis for the possible quantum violations for these non-contextual inequalities and find that our graphical KS inequality approach may help us to improve some state-independent proofs for the Kochen-Specker theorem in our anther work\cite{Tang-yu}.

\section{Description of the $(6n+2)$-ray model}

Conventionally,  the notation of a ray(normalized unless emphasized) in the topic of quantum contextuality is more commonly used than its two  alternatives: a complex vector in the Hilbert space and a normalized rank-1 projector on the vector. To be specific, a ray $|{\psi}_i\rangle=\alpha_i|0\rangle+\beta_i|1\rangle+\gamma_i|2\rangle$ ($\alpha_i,\beta_i,\gamma_i\in \mC$) can represent $\textbf{r}_i=(\alpha_i,\beta_i,\gamma_i)$ or or $P=|{\psi}_i\rangle\langle{\psi}_i|$. Accordingly, the orthogonality and normalization, $\langle\psi_i|{\psi}_j\rangle=\delta_{ij}$,  can be written as $\textbf{r}_i^{\ast}\textbf{r}_j=\delta_{ij}$.

\begin{figure}
\includegraphics[scale=0.7]{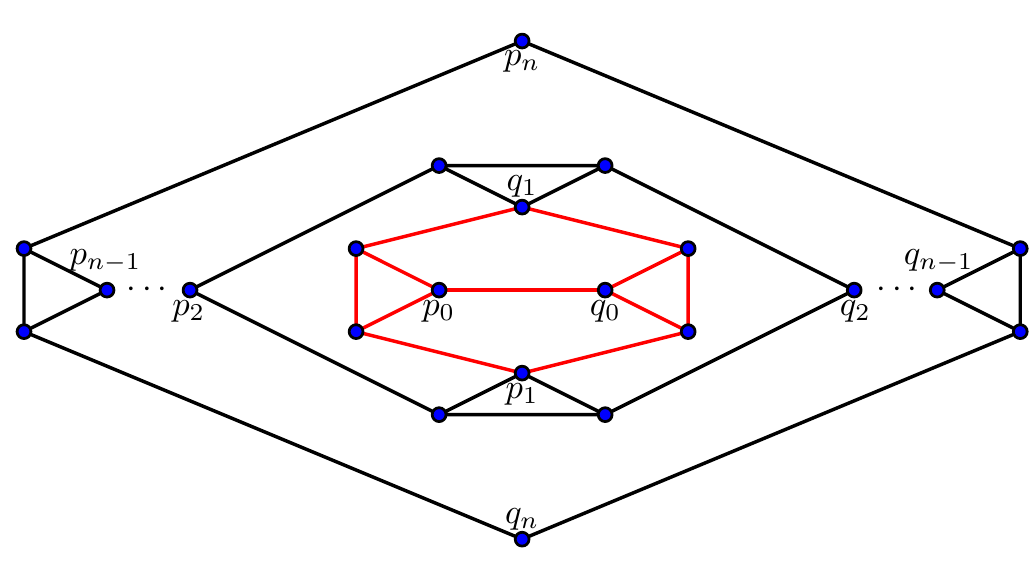} \caption{\label{6nplus2} Graphical representation for the $(6n+2)$-ray model. The blue dots stand for the rays and the links indicate the orthogonality for the connected rays. Note that the center 8 rays linked with red line generate the  Clifton's 8-ray model.}
\end{figure}

For any two rays $P_{\phi}=|\phi\rangle\langle\phi|$ and $P_{\psi}=|\psi\rangle\langle\psi|$, if $|\langle\psi|\phi\rangle|\le{\frac{n}{n+2}}$, we can always add $2n$ complete orthonormal bases to build a $(6n+2)$-ray model\cite{Tang-yu}.  The graphical representation for this model is shown in FIG.\ref{6nplus2}, where $p_n$ and $q_n$ stand for the rays $P_{\phi}$ and $P_{\psi}$ respectively. And one can easily see the orthogonal relations for all these rays from this graph. Clearly, this model can be considered as a generalization of the Clifton's 8-ray model\cite{Clifton} as the latter is exactly the case for $n=1$.

Analogous to the Clifton's 8-ray model, if we assign value 1 to two rays $p_n$ and $q_n$ simultaneously by the non-contextual hidden variable theory, it is not difficult for us to get a contradiction that two orthogonal rays $p_0$ and $q_0$ should also be assigned to value 1. This is why the $(6n+2)$-ray model can be considered as a proof for quantum contextuality, although in a state-dependent manner.

We can also get a non-contextual inequality  from the $(6n+2)$-ray model. For some systems with complicated algebraic structures, a regular method to get the upper bounds for the non-contextual inequalities is by the computer search\cite{cabello08}.
Though the $(6n+2)$-ray model is somewhat complicated,  we have already derived the upper bound in Ref.\cite{Tang-yu} by a exact algebraic approach rather than by a computer search.
Next, we give a brief introduction to this approach.

\begin{figure}
\includegraphics[scale=0.7]{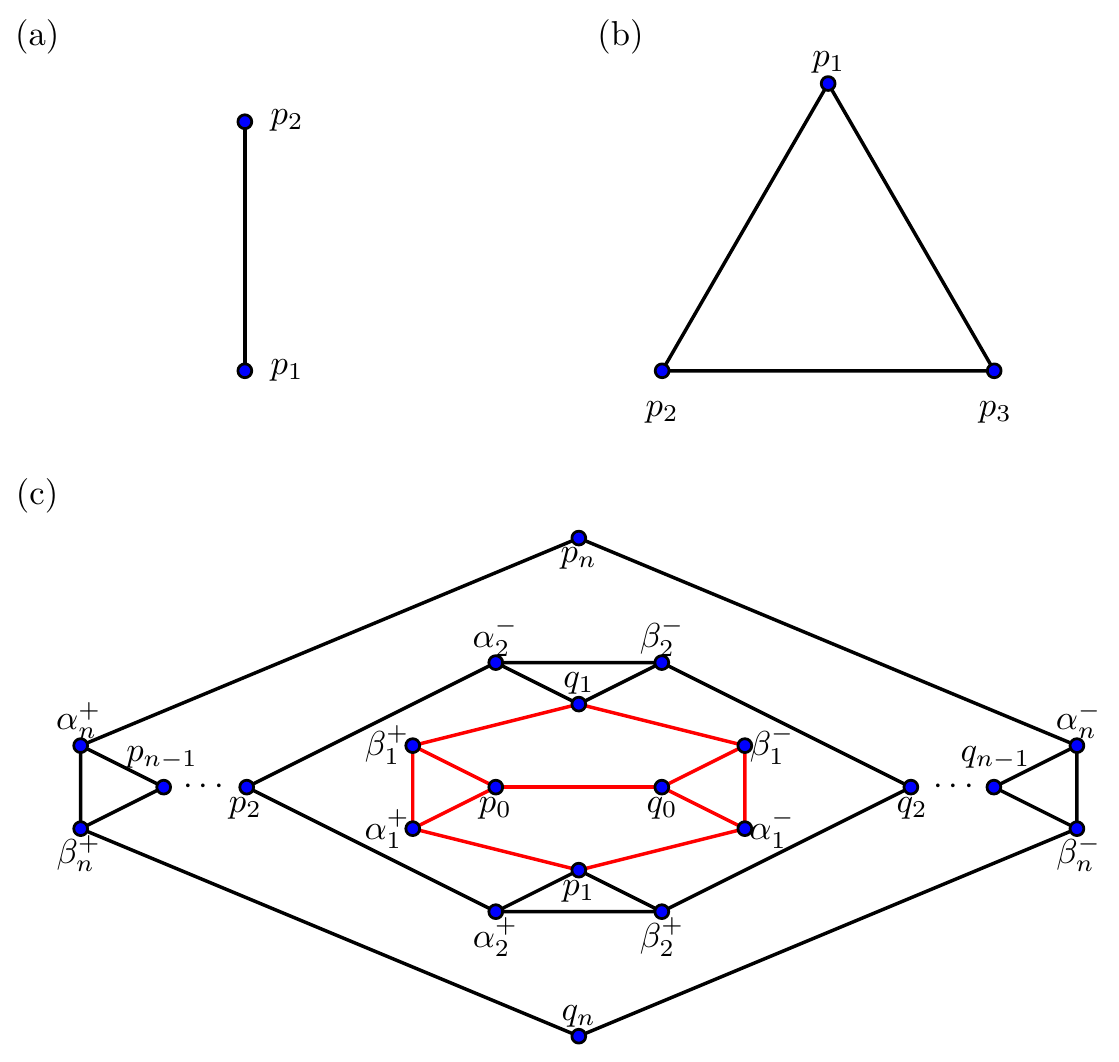} \caption{\label{ori-model} Each graph can give a value assignment inequality,  where the vertices represent the rays and the edges show the orthogonality for the rays. (a)Two orthogonal rays. (b)A complete orthonormal base for $d=3$.  (c) The $(6n+2)$-ray model.}
\end{figure}

First, it is clear that the case for the value assignment to a single ray or several independent (unconnected) rays is trivial. Thus the two-ray value assignment inequality from  FIG.\ref{ori-model}-(a)  can be considered as the simplest and nontrivial one. In spite of  failure to  show the  quantum contextuality by the inequality itself, it is quite useful in constructing  more  complicated  non-contextual inequalities.

We denote by each index of the vertex in FIG.\ref{ori-model} the related ray. Then value 0 or 1 can be assigned to each of them, and  an inequality  for FIG.\ref{ori-model}-(a) can be given by
\begin{align}\label{2rayineq}
   \langle\mA_2\rangle=p_1+p_2-p_1p_2\leq1.
\end{align}
The proof is straightforward.  Let $\bar{p}_i=1-p_i\in\{0,1\},~i=1,2$,  then we have
$\langle\mA_2\rangle=1-\bar{p}_1\bar{p}_2\leq1$.

Note that we have omit the bra and ket notations in the expansion of $\langle\mA_2\rangle$ as it will not cause any confusion in the classical case. Hereafter we may follow the same convention for simplicity.

For FIG.\ref{ori-model}-(b), the value assignment inequality is
\begin{align}\label{tri}
   \langle\mA_{\bigtriangleup}\rangle=\sum_{i=1}^3p_i-\sum_{i<j\in\{1,2,3\}}p_ip_j\leq1.
\end{align}
We can get the value assignment upper bound for $\langle\mA_{\bigtriangleup}\rangle$ with the help of $\langle\mA_2\rangle$.  For $\sum_{i=1}^3p_i\leq1$, we can see clearly that $\langle\mA_{\bigtriangleup}\rangle\leq1$ from its expansion in Eq.(\ref{tri}), and for $\sum_{i=1}^3p_i\geq2$, we can also get the same conclusion
from $\langle\mA_{\bigtriangleup}\rangle=3\langle\mA_2\rangle-\sum_{i=1}^3p_i\leq3-2=1$.

Note that here $3\langle\mA_2\rangle\equiv\langle\mA_2\rangle_1+\langle\mA_2\rangle_2+\langle\mA_2\rangle_3$, where $\langle\mA_2\rangle_i=p_i+p_{i+1}-p_ip_{i+1},~(i=1,2,3)$ and $p_4\equiv p_1$. In what follows, the similar notations will be used unless specified.

Before discussing the KS value assignment inequality for the $(6n+2)$-ray model, we would like to introduce a special observable, which will be considered as a ``hyper-edge" operator in the following text and can be defined as
\begin{align}\label{hyperedge1}
    C(p_n,q_n)=\sum_{i\in\mV}v_i-\sum_{i<j,(i,j)\in\mE}v_iv_j-p_n-q_n,
\end{align}
where $\mV$ is the index set for all the vertices of the graph in FIG.\ref{ori-model}-(c) and $\mE$ is an index representation for  the edge set. In other words, $(i,j)\in\mE$ indicates that $(v_i,v_j)$ is in the edge set of the graph. Then we can get the following value assignment inequality
\begin{align}\label{hyperedgeineq}
    \langle C(p_n,q_n)\rangle\leq2n.
\end{align}
This can be derived from another form of $\langle C(p_n,q_n)\rangle$, namely,
\begin{align}\label{hyperedge2}
    \langle C(p_n,q_n)\rangle=&2n\langle\mA_{\bigtriangleup}\rangle
    -\sum_{i=1}^np_i(\alpha_i^++\alpha_i^-)\cr
    &-\sum_{i=1}^nq_{i}(\beta_i^++\beta_i^-)-p_0q_0\cr
    \leq2n.
\end{align}

Let us return to the $(6n+2)$-ray model in FIG.\ref{ori-model}-(c).

{\it Lemma. ---} The KS value assignment inequality for the $(6n+2)$-ray model can be given by
\begin{align}\label{extendedineq}
    \langle \mB_n\rangle=\sum_{i\in\mV}v_i-\sum_{i<j,(i,j)\in\mE}v_iv_j\leq&1+2n.
\end{align}

{\it Proof.---} Here we give a proof which is different from the approach in Ref.\cite{Tang-yu}.

First, Eq.(\ref{extendedineq}) holds for $n=0$ since $\langle \mB_0\rangle=\langle\mA_2\rangle$.

Assume that the statement is also true for $n-1~(n\geq1)$, namely, $\langle\mB_{n-1}\rangle\leq2n-1$. Then we we should prove that it holds for $n$.
Notice that $\langle \mB_n\rangle$ can also be written as
\begin{align}\label{extendedineqnew1}
    \langle\mB_n\rangle=&\langle\mB_{n-1}\rangle+\alpha_n^++\alpha_n^-+\beta_n^++\beta_n^-+p_n+q_n\cr
    &-p_{n-1}(\alpha_n^++\beta_n^+)-q_{n-1}(\alpha_n^-+\beta_n^-)\cr
    &-p_{n}(\alpha_n^++\alpha_n^-)
    -q_{n}(\beta_n^++\beta_n^-)\cr
    &-\alpha_n^+\beta_n^+-\alpha_n^-\beta_n^-,
 \end{align}
or
\begin{align}\label{extendedineqnew2}
     \langle\mB_n\rangle=&\langle\mB_{n-1}\rangle+2\langle\mA_{\bigtriangleup}\rangle+4\langle\mA_2\rangle\cr
     &-(\alpha_n^++\alpha_n^-+\beta_n^++\beta_n^-+p_n+q_n)\cr
     &-(p_{n-1}+q_{n-1})\cr
     =&\langle C(p_{n-1},q_{n-1})\rangle+2\langle\mA_{\bigtriangleup}\rangle+4\langle\mA_2\rangle\cr
     &-(\alpha_n^++\alpha_n^-+\beta_n^++\beta_n^-+p_n+q_n).
\end{align}

(i)For the case of $\alpha_n^++\alpha_n^-+\beta_n^++\beta_n^-+p_n+q_n\leq2$,  we have $\langle\mB_n\rangle\leq\langle\mB_{n-1}\rangle+2\leq2n+1$ by Eq.(\ref{extendedineqnew1});
(ii)and when $\alpha_n^++\alpha_n^-+\beta_n^++\beta_n^-+p_n+q_n\geq3$, according to Eq.(\ref{extendedineqnew2}), we can get $\langle\mB_n\rangle\leq2(n-1)+2+4-3=2n+1$.

Therefore, $\langle\mB_n\rangle\leq2n+1$ holds for any non-negative integer $n$.\hfill$\sharp$

Next, some definitions from graph theory should be given before discussing our main results.

\section{An introduction to the hyper-graphs}

It is known that one of the original constraints for the KS value assignment requires that two mutually orthogonal rays cannot be assigned to value 1 simultaneously. This constraint can be generalized to two ordinary rays by the $(2n+6)$-ray model. To be specific, if two rays $|\phi\rangle$ and $|\psi\rangle$ satisfy $|\langle\psi|\phi\rangle|\le{\frac{n}{n+2}}$, they can always generate a $(2n+6)$-ray model by adding $2n$ auxiliary complete orthonormal bases,
such that $|\phi\rangle$ and $|\psi\rangle$ can not be both assigned to value 1. This motivates us to defined a new graphical structure to enrich the  original graphical representation.
Later we will see that this  structure will facilitate us to analytically derive the upper bounds for various non-contextual inequalities.

 \begin{figure}
\includegraphics[scale=0.5]{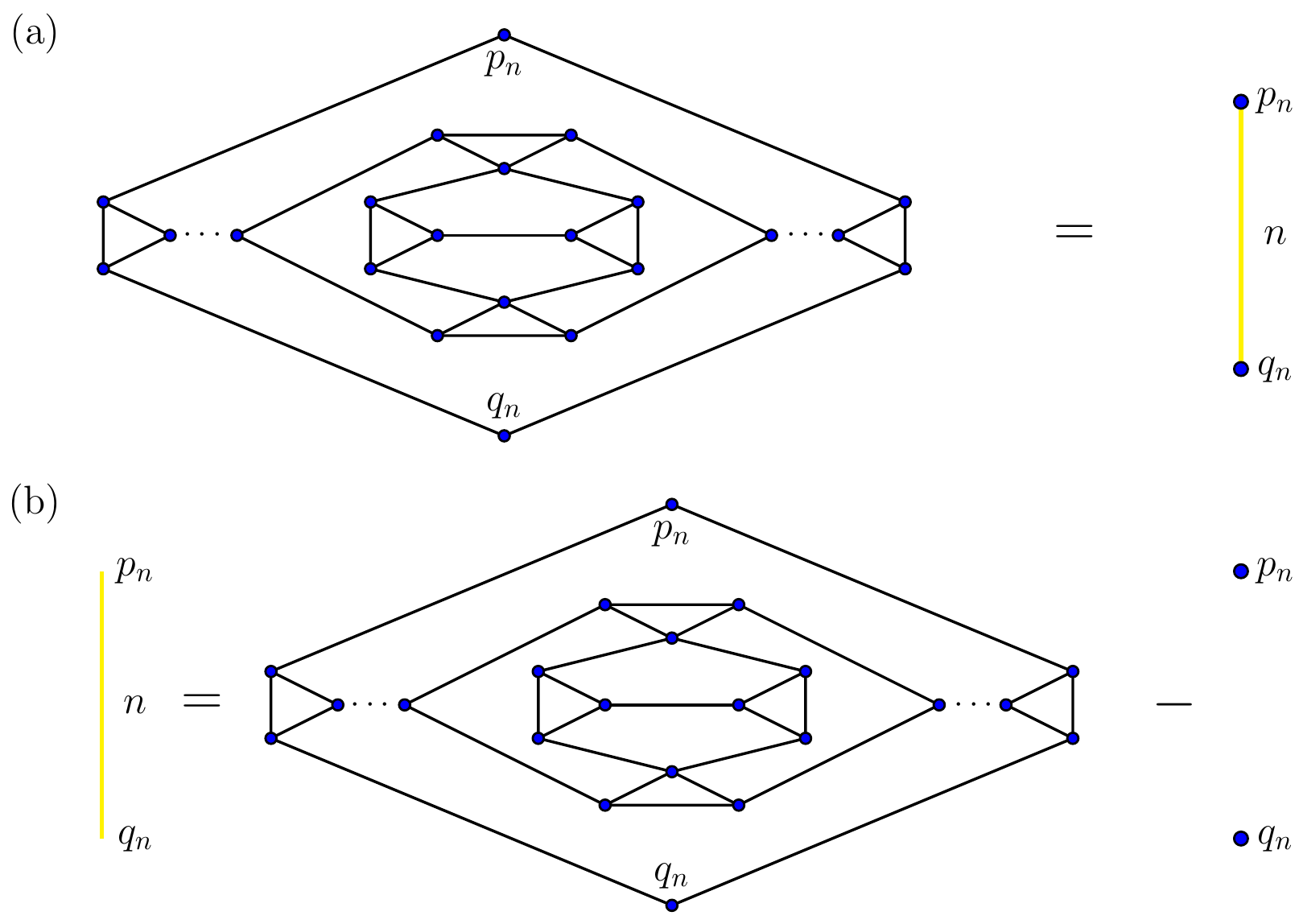} \caption{\label{hyper-edge} (a)The hyper-graph  associating with the $(6n+2)$-ray model. (b)Graphical representation for a hyper-edge.}
\end{figure}

Notice that we have already presented a systematic and programmable approach to
construct a state-independent  proof of the KS theorem  for the first time in Ref.\cite{Tang-yu} based on the $(6n+2)$-ray model. Actually, a state-dependent proof which seems much simpler, can also be constructed by the same method via reducing some constraints.
Next we give a brief review on this approach.

First, we choose several nonspecific (usually nonparallel or nonorthogonal) rays as a
fundamental ray set $\mF=\{|\psi_i\rangle\}_{i\in I}$ (or $\mF=\{P_i|P_i=|\psi_i\rangle\langle\psi_i|\}_{i\in I}$), where $I$ is an index set and the number of rays in $\mF$ is $|I|$. Then for a fixed $N~(N>0,N\in\mZ)$, considering any two rays $|\psi_k\rangle$ and $|\psi_l\rangle$ from $\mF$, if they satisfy $|\langle\psi_k|{\psi}_l\rangle|\in(\frac{n-1}{n+1},\frac{n}{n+2}]~(0<n\leq N,n\in\mZ)$, we can economically build a $(6n+2)$-ray model by adding $2n$ extra complete orthonormal bases. Take FIG.\ref{6nplus2} or FIG.\ref{ori-model}-(c) for example, what we need to do is just to replace $p_n$ and $q_n$ with $P_k=|\psi_k\rangle\langle\psi_k|$ and $P_l=|\psi_l\rangle\langle\psi_l|$ respectively. And an $n$-weighted {\it{hyper-edge}} linking the two rays  can be defined as all the rays from the complete orthonormal bases together with all the edges from the original graphical representation for this $(6n+2)$-ray model, see FIG.\ref{hyper-edge}-(b), where each orange line represents a hyper-edge.  Repeat this operation to other pairs of rays in $\mF$, and we can construct a proof for the KS theorem and get the corresponding hyper-graph $G$. Clearly, the simplest nontrivial  hyper-graph is exactly the representation for the $(6n+2)$-ray model(FIG.\ref{hyper-edge}-(a)).

We denote by $V$ and $E$  the vertex set and the hyper-edge set of a hyper-graph $G$ respectively. Without loss of generality, we denote $V$ as $V=\mF=\{P_i|i=1,2,...,|V|\}$ and we have $|V|=|I|$. Note that a $0$-weighted hyper-edge between two vertices is an edge of a normal graph. And do not confuse two unconnected vertices with a two-vertex hyper-graph whose hyper-edge is $0$-weighted. From this point of view, a normal graph is only a special case of a hyper-graph. This is why we use the same notation $G$ to denote them for simplicity.

For each hyper-graph $G$,  we can associate with the following KS observable with respect to the non-contextual inequality\cite{Tang-yu}
\begin{equation}
G=\sum_{i=1}^{|V|}P_i+\sum_{i=1}^{|V|-1}\sum_{j\in \mN_i,j>i} C(P_i,P_j),
\end{equation}
where $C(P_i,P_j)$  defined by Eq.(\ref{hyperedge1}) can be referred to as a {\it hyper-edge observable}, and $\mN_i$ stands for index set for the neighborhood of the vertex $P_i$, i.e., if $j\in\mN_i$, then $(P_i,P_j)\in E$. For our optimal construction of a proof for the  KS theorem (by adding the minimum number of complete orthonormal bases between any two rays in $V$),  $C(P_i,P_j)$ vanishes if $|\langle \psi_i|\psi_j\rangle|>\frac{N}{N+2}$ and involves $2n_{ij}$ complete orthonormal bases where
$$n_{ij}=\left\lceil\frac{2|\langle \psi_i|\psi_j\rangle|}{1-|\langle \psi_i|\psi_j\rangle|}\right\rceil.$$
From Eq.(\ref{hyperedgeineq}), we can get $\langle C(P_i,P_j)\rangle\le 2n_{ij}$. But for a non-optimal construction, the number of the orthonormal bases corresponding to $C(P_i,P_j)$ is  usually larger than $n_{ij}$ and $C(P_i,P_j)$ vanishes if $(P_i,P_j)\notin E$.
In what follows, we only care about the non-contextual inequality from a given hyper-graph rather than the construction of a proof for KS theorem. Hence other problems such as the optimization for $n_{ij}$ will be ignored.

\begin{figure}
\includegraphics[scale=0.6]{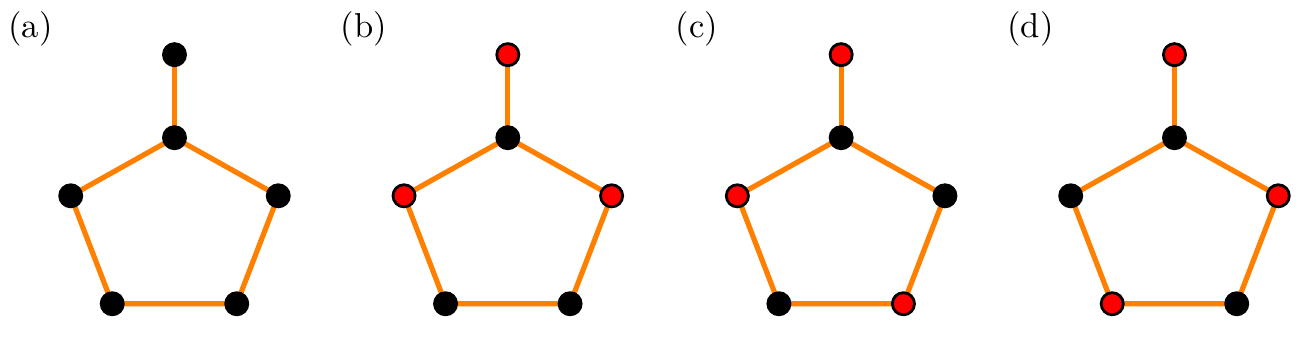} \caption{\label{MUVS} (a)A six-vertex hyper-graph. (b)-(d)Vertices colored with red generate three different maximal unconnected vertex sets of this hyper-graph. }
\end{figure}

A vertex set $U$ $(U\subset V)$  is called a {\it maximal unconnected vertex set} of a hyper-graph $G$ if, (i) for any two vertices $P_i$ and $P_j$ in $U$, $(P_i,P_j)\notin E$; (ii) for any other set $U^{\prime}\subset V$ satisfies (i), the vertex number $|U^{\prime}|\leq|U|$. Usually, this set is not unique. Such an example is given in FIG.\ref{MUVS}.

\begin{figure}
\includegraphics[scale=0.6]{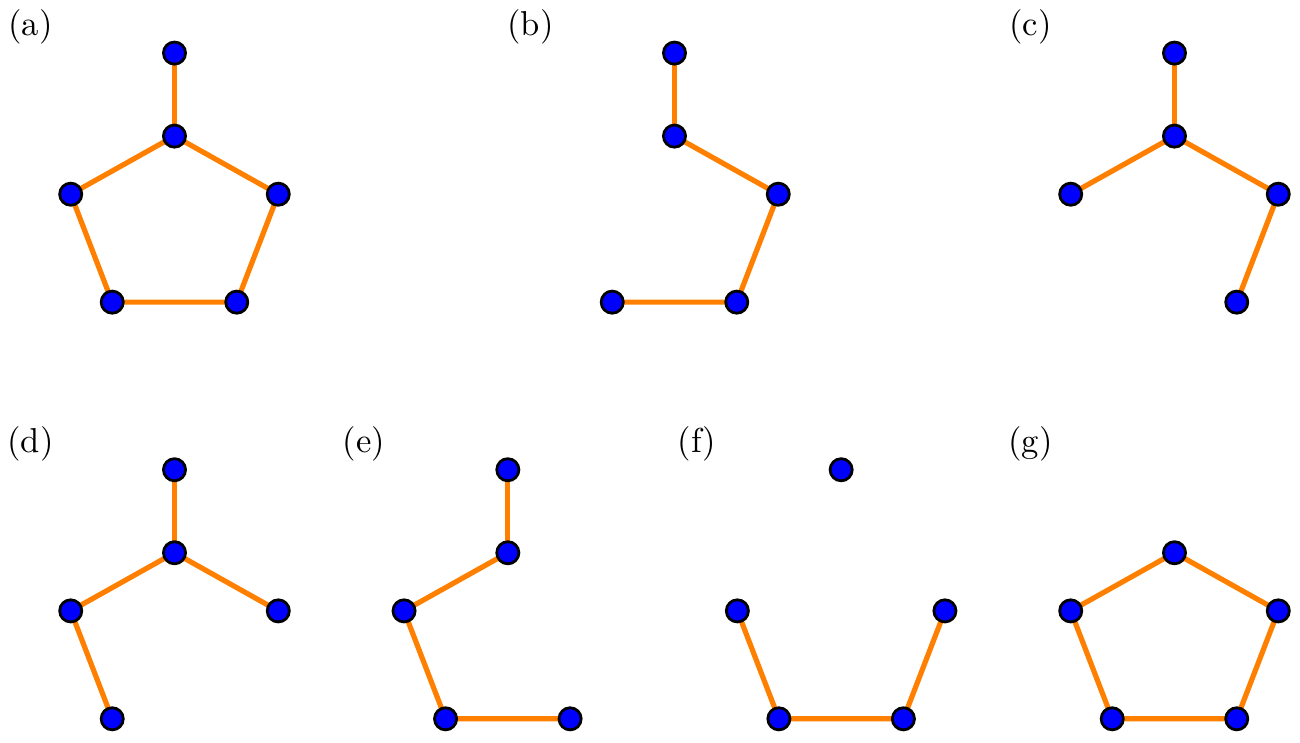} \caption{\label{subgraphdecomp} (a)A six-vertex hyper-graph. (b)-(g)Six subgraphs by removing one vertex from the hyper-graph in (a). }
\end{figure}

Let us return to the case of normal graphs. A {\it subgraph} $G^{\prime}$ of a graph $G$ is also a graph whose vertex set satisfies $V^{\prime}\subset V$ and whose edge set $E^{\prime}$ consists of all of the edges in $E$ that have both endpoints in $V^{\prime}$. But here $V$ and $E$ only stand for the vertex set and the edge set of the normal graph $G$, which is different from the notations referred above.
This definition can be easily generalized to the hyper-graph case replacing the edge  with hyper-edge. And we are supposed to  call it ``sub-hyper-graph", but for convenience we would still refer to it as ``subgraph".  FIG.\ref{subgraphdecomp} gives us an example for all the five-vertex subgraphs from a six-vertex hyper-graph.

If we denote by $G^i$ the subgraph obtained by removing the vertex $P_i$ and all the edges with one of the endpoint $P_i$ in $G$, then it holds
\begin{equation}
\langle G\rangle=\langle G^i\rangle+\langle P_i\rangle+\sum_{j\in \mN_i}\langle  C(P_i,P_j)\rangle.
\end{equation}
And we can get
\begin{align}
|V|\langle G\rangle&=\sum_{i=1}^{|V|}\langle G^i\rangle+\sum_{i=1}^{|V|}\langle P_i\rangle+\sum_{i=1}^{|V|}\sum_{j\in\mN_i}\langle  C(P_i,P_j)\rangle\cr
&=\sum_{i=1}^{|V|}\langle G^i\rangle-\sum_{i=1}^{|V|}\langle P_i\rangle+2\langle G\rangle
\end{align}
or a more compact form\cite{Tang-yu}
\begin{eqnarray}\label{subg}
(|V|-2)\langle G\rangle&=&{\sum_{i=1}^{|V|}\langle G^i\rangle-\sum_{i=1}^{|V|}\langle P_i\rangle}.
\end{eqnarray}
This can be considered as a relation of the subgraph decomposition.

\section{ Three typical non-contextual inequalities}

\begin{figure}
\includegraphics[scale=0.65]{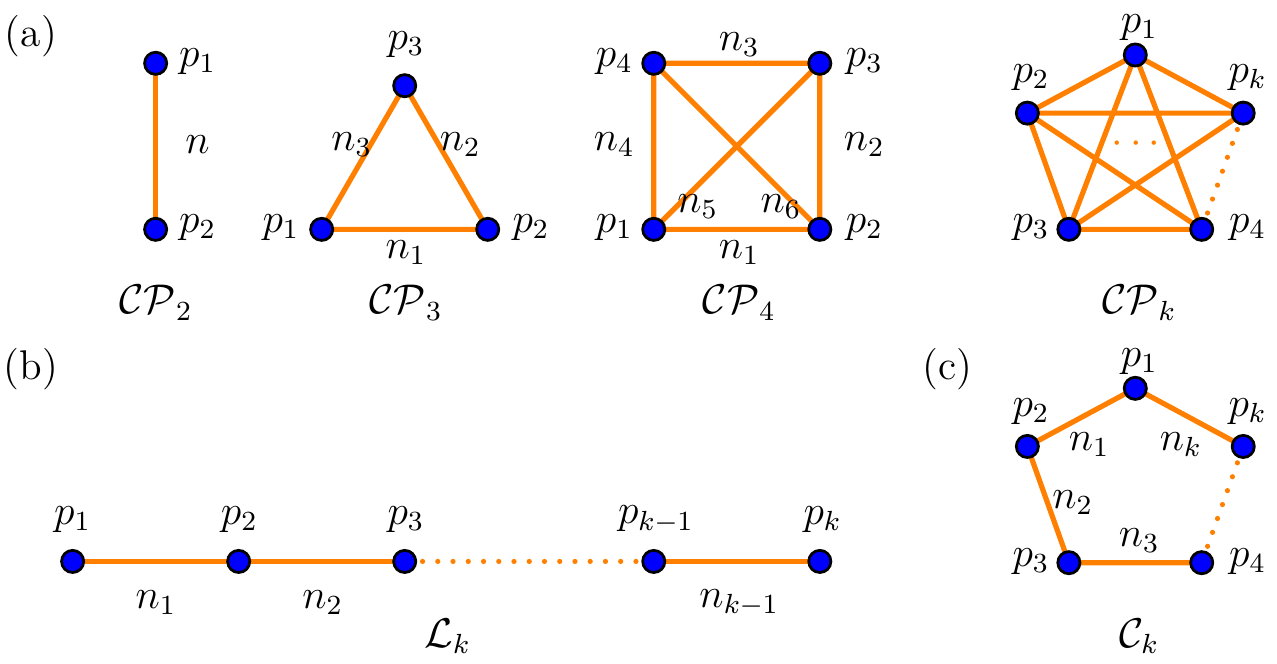} \caption{\label{3-special-graph} (a)The family of complete hyper-graphs $\{\mCP_i,i=2,3,...,k\}$. (b) A $k$-vertex linear hyper-graph $\mL_k$ (c)A $k$-vertex cyclic hyper-graph $\mmC_k$.}
\end{figure}

Let us consider some typical hyper-graphical structures and the related non-contextual inequalities. Here we mainly discuss three different families of hyper-graphs. Based on the hyper-graphical representation for the $(6n+2)$-ray model or Lemma, i.e., $\langle\mCP_2\rangle=\langle\mL_2\rangle=\langle\mB_n\rangle\leq2n+1$ (see FIG.\ref{3-special-graph}), we can derive the non-contextual inequalities analytically from the hyper-graphs with more vertices in FIG.\ref{3-special-graph}.

\subsection{Complete hyper-graphs}

Analogous to the definition of a complete graph in graph theory, a complete hyper-graph is a hyper-graph in which every pair of  vertices
is connected  by a hyper-edge. Then we can get the following theorem.

{\it Theorem 1.---} The non-contextual inequality associated with  a $k$-vertex complete hyper-graph ($k\geq2$) can be written as
\begin{align}\label{CPgraph}
    \langle\mCP_k\rangle=\sum_{i=1}^kp_i+\sum_{i=1}^{k-1}\sum_{j=2,j>i}^k\langle C(p_i,p_j)\rangle\leq2\sum_{i=1}^{\frac{k(k-1)}{2}}n_i+1,
\end{align}
where $p_i$ is the $i$-th ray (vertex) and $n_i$ stands for the weight for the $i$-th hyper-edge in FIG.\ref{3-special-graph}-(a).

{\it Proof.---} Clearly the statement is true for $k=2$.

Assume that  Eq.(\ref{CPgraph}) holds for any $\langle\mCP_{k-1}\rangle$ $(k\geq3)$. 
We should prove that  Eq.(\ref{CPgraph}) also holds for $\langle\mCP_{k}\rangle$.
From Eq.(\ref{subg}), we have
$$(k-2)\langle\mCP_k\rangle=\sum_{i=1}^k\langle\mCP_k^i\rangle-\sum_{i=1}^{k}p_i$$.

(i)If $\sum_{i=1}^{k}p_i\leq1$, we can see that  Eq.(\ref{CPgraph}) still holds from Eq.(\ref{hyperedgeineq}) and the expansion for $\langle\mCP_k\rangle$ in Eq.(\ref{CPgraph}).

(ii)If $\sum_{i=1}^{k}p_i\geq2$, from another form of $\langle\mCP_k\rangle$ referred above, we have
$$(k-2)\langle\mCP_k\rangle\leq k+(k-2)\cdot2\sum_{i=1}^{\frac{k(k-1)}{2}}n_i-2,$$
namely,
$$\langle\mCP_k\rangle\leq2\sum_{i=1}^{\frac{k(k-1)}{2}}n_i+1.$$

Therefore, Eq.(\ref{CPgraph}) holds for any $\langle\mCP_{k}\rangle$ $(k\geq2)$. \hfill$\sharp$

\subsection{Linear hyper-graphs}

Likewise, for the linear hyper-graph shown in FIG.\ref{3-special-graph}-(b), we have the following theorem.

{\it Theorem 2.---} For a $k$-vertex linear hyper-graph, the non-contextual inequality
can be given by
\begin{align}\label{Lgraph}
    \langle\mL_k\rangle=\sum_{i=1}^kp_i+\sum_{i=1}^{k-1}\langle C(p_i,p_{i+1})\rangle\leq2\sum_{i=1}^{k-1}n_i+\lceil\frac{k}{2}\rceil,
\end{align}
where $n_i$ is the weight for the relevant hyper-edge in FIG.\ref{3-special-graph}-(b).

{\it Proof.---}
Clearly, Eq.(\ref{Lgraph}) holds for $k=2$. 
Assume that it is also true for $\langle\mL_{k-1}\rangle$ $(k\geq3)$. Next let us prove that Eq.(\ref{Lgraph})  holds for $\langle\mL_{k}\rangle$.

(i)From the expansion of $\langle\mL_k\rangle$ in Eq.(\ref{Lgraph}), it is clear that the inequality holds for the case $\sum_{i=1}^{k}p_i\leq\lceil\frac{k}{2}\rceil$.

(ii)If $\sum_{i=1}^{k}p_i\geq\lceil\frac{k}{2}\rceil+1$, we can use another form of
 $\langle\mL_k\rangle$, which reads
 $$\langle\mL_k\rangle=(k-1)\langle\mCP_2\rangle+p_1+p_k-\sum_{i=1}^{k}p_i.$$ 
Therefore, 
\begin{align*}
 \langle\mL_k\rangle\leq& 2\sum_{i=1}^{k-1}n_i+k-1+2-\sum_{i=1}^{k}p_i\cr
  \leq&2\sum_{i=1}^{k-1}n_i+k-\lceil\frac{k}{2}\rceil\cr
 \leq&2\sum_{i=1}^{k-1}n_i+\lceil\frac{k}{2}\rceil.
\end{align*}

Hence Theorem 2 holds for all possible KS value assignments to the related rays. \hfill$\sharp$

\subsection{Cyclic hyper-graphs}

Another  non-contextual inequality from the cyclic hyper-graph in FIG.\ref{3-special-graph}-(c) is shown in below.

{\it Theorem 3.---} For a $k$-vertex cyclic hyper-graph, the non-contextual inequality
can be given by
\begin{align}\label{Cgraph}
    \langle\mmC_k\rangle=\sum_{i=1}^kp_i+\sum_{i=1}^{k}\langle C(p_i,p_{i+1})\rangle\leq2\sum_{i=1}^{k}n_i+\lfloor\frac{k}{2}\rfloor.
\end{align}
where $n_i$ represents the weight for the hyper-edge linking the rays $p_i$ and $p_{i+1}$, and $p_{k+1}\equiv p_{1}$.

{\it Proof.---}
Similar to the proof in Theorem 2, we can also write $\langle\mmC_k\rangle$ in another form
$$\langle\mmC_k\rangle=k\langle\mCP_2\rangle-\sum_{i=1}^{k}p_i$$.

(i)By the original expression for $\langle\mmC_k\rangle$ in Eq.(\ref{Cgraph}), it is clear that the inequality holds for the case $\sum_{i=1}^{k}p_i\leq\lfloor\frac{k}{2}\rfloor$.

(ii)If $\sum_{i=1}^{k}p_i\geq\lfloor\frac{k}{2}\rfloor+1$, we can use the above second form of
 $\langle\mmC_k\rangle$. That is
\begin{align*}
 \langle\mmC_k\rangle\leq& 2\sum_{i=1}^{k}n_i+k-\sum_{i=1}^{k}p_i\cr
  \leq&2\sum_{i=1}^{k}n_i+k-1-\lfloor\frac{k}{2}\rfloor\cr
 \leq&2\sum_{i=1}^{k}n_i+\lfloor\frac{k}{2}\rfloor.
\end{align*}

Therefore, Theorem 3 holds for any KS value assignment. \hfill$\sharp$

\section{Non-contextual inequality for an ordinary hyper-graph}

For any ordinary hyper-graph, we give a theorem which is equivalent to a conclusion in Ref.\cite{Tang-yu} to describe the non-contextual inequality.

{\it Theorem 4.---} For a $k$-vertex  hyper-graph $G_k$,  we can always find at least one maximal unconnected vertex set $U$. If we denote by $p_i$ and $n_j$ the $i$-th vertex and the weight for the $j$-th hyper-edge, then we have the following non-contextual inequality
\begin{align}\label{Rgraph}
    \langle G_k\rangle=\sum_{i=1}^kp_i+\sum_{i=1}^{k-1}\sum_{j\in\mN_i,j>i}\langle C(p_i,p_j)\rangle\leq2\sum_{i=1}^{|E|}n_i+|U|,
\end{align}
where $E$ is the hyper-edge set.

{\it Proof.---}Equivalently, we can prove it by verifying another proposition. That is, if such an inequality can be proved to be true for all the possible hyper-graphs with a fixed maximal unconnected vertex set $U$ ($|U|<k$), then it also holds for a hyper-graph with $k$ vertices.

We denote by $V$ the vertex set of the hyper-graph $G_{|V|}$, and label the vertices in $U$ by $p_1,p2,...,p_{|U|}$.

The case for $|V|=|U|$ is trivial.

For $|V|=|U|+1$, at least one hyper-edge with endpoints $p_{|U|+1}$ and some vertex in $U$ can be found by the definition of the maximal unconnected vertex set. If $E=\{(p_{|U|+1},p_{\alpha_1}),(p_{|U|+1},p_{\alpha_2}),...,(p_{|U|+1},p_{\alpha_l})\}$. Then we have
\begin{align*}
\langle G_{|U|+1}\rangle=&\sum_{i=1,i\neq\alpha_1}^{|U|}p_i+\langle\mCP_2\rangle_{p_{|U|+1}p_{\alpha_1}}\cr
&+\sum_{i=2}^{l}\langle C(p_{|U|+1},p_{\alpha_i})\rangle\cr
\leq&|U|-1+2n_1+1+2\sum_{i=2}^{|E|}n_i\cr
=&2\sum_{i=1}^{|E|}n_i+|U|.
\end{align*}

Assume that  Eq.(\ref{Rgraph}) holds for all the hyper-graphs with $|V|=|U|+m~(m>1)$.
Then for $|V|=|U|+m+1$, if $\sum_{i=1}^{|U|+m+1}p_i\leq|U|$, it is clear that Eq.(\ref{Rgraph}) is true. Therefore, we only need to check the case for $\sum_{i=1}^{|U|+m+1}p_i\geq|U|+1$.
From the subgraph decomposition relation Eq.(\ref{subg}), we have
\begin{align*}
&[(|U|+m+1)-2]\langle G_{|U|+m+1}\rangle\\
=&\sum_{\alpha=1}^{|U|+m+1}\langle G_{|U|+m}^{\alpha}\rangle-\sum_{i=1}^{|U|+m+1}p_i\\
\leq&{(|U|+m+1)}|U|+2(|U|+m+1-2)\sum_{i=1}^{|E|}n_i\\
&-(|U|+1)\\
=&{(|U|+m-1)}(|U|+2\sum_{i=1}^{|E|}n_i)+(|U|-1).
\end{align*}
Thus $\langle G_{|U|+m+1}\rangle\leq2\sum_{i=1}^{|E|}n_i+|U|+\frac{|U|-1}{|U|+m-1}$. Since
for any value assignment, $\langle G_{|U|+m+1}\rangle$ should be an integer, and $\frac{|U|-1}{|U|+m-1}<1$, then $\langle G_{|U|+m+1}\rangle\leq2\sum_{i=1}^{|E|}n_i+|U|$.
Hence the  Eq.(\ref{Rgraph}) holds for any hyper-graph with a maximal unconnected vertex set $U$.  As a special case, it also holds for $G_k$.\hfill$\sharp$

By Theorem 4, one can also derive the Eqs.(\ref{CPgraph},\ref{Lgraph},\ref{Cgraph}) by counting the number of the vertices in their maximal unconnected vertex sets.

\section{Non-contextual inequalities for some fractal structures}

\begin{figure}
\includegraphics[scale=0.65]{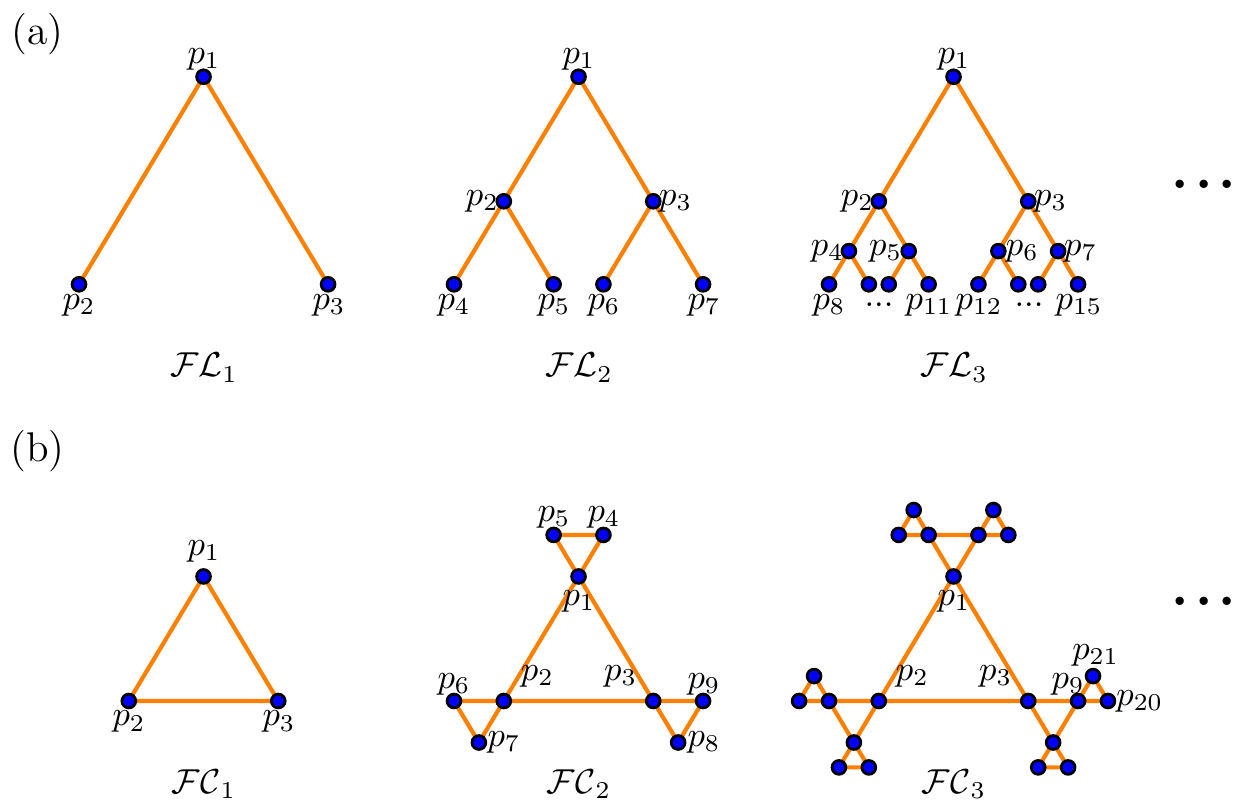} \caption{\label{frac} (a)A  fractal tree hyper-graph family $\{\mFL_i,i=2,3,...,k\}$. (b) A fractal cyclic  hyper-graphs family $\{\mFC_i,i=2,3,...,k\}$.}
\end{figure}

If the vertex number of an ordinary hyper-graph is large, then for any KS value assignment, the upper bound for its non-contextual inequality is very difficult to calculate. But in some special cases, analytical formulas for the upper bounds can be recursively derived. We have already given three families of such examples in previous sections.
Here we present  two more examples, which come from the fractal hyper-graphs.

Considering a fractal hyper-graph family, e.g. FIG.\ref{frac}-(a), it is not difficult for us to notice that some former methods to derive the upper bound  of a non-contextual  inequality  may not work effectively in this scenario, e.g., the way used in Theorem 2 and Theorem 3. But fortunately another approach by Theorem 4 seems to be a nice choice. As for some fractal hyper-graph structures, it is easy to find out their maximal unconnected vertex sets.

For the fractal hyper-graph families $\{\mFL_k\}_{k\in{\mathbb{N}}}$ and $\{\mFC_k\}_{k\in{\mathbb{N}}}$  FIG.\ref{frac}, if we denote by $U_{\mFL}^k$ $(U_{\mFC}^k)$ the maximal unconnected vertex set for the $k$-th graph in $\{\mFL_k\}_{k\in{\mathbb{N}}}$ $(\{\mFC_k\}_{k\in{\mathbb{N}}})$, then
\begin{align*}
    |U_{\mFL}^k|=&2^k\cdot\frac{1-(\frac{1}{4})^{\lfloor\frac{k}{2}\rfloor+1}}{1-\frac{1}{4}}
    =\frac{4}{3}\cdot(2^k-2^{k\mod2-2});\cr
    |U_{\mFC}^k|=&2^k-1.
\end{align*}
Therefore, we have
\begin{align}\label{FLgraph}
    \langle \mFL_k\rangle=&\sum_{i=1}^{2^{k+1}-1}p_i+\sum_{i=1}^{2^{k}-1}(\langle C(p_i,p_{2i})\rangle+\langle C(p_i,p_{2i+1}))\cr
    \leq&2\sum_{i=1}^{2^{k+1}-2}n_i+|U_{\mFL}^k|,
\end{align}
and
\begin{align}\label{FLgraph}
    &\langle \mFC_k\rangle\cr
    =&\sum_{i=1}^{3\cdot(2^k-1)}p_i+\langle C(p_1,p_{2})\rangle+\langle C(p_1,p_{3})\rangle+\langle C(p_{2},p_{3})\rangle\cr
    &+\sum_{i=1}^{3\cdot(2^{k-1}-1)}(\langle C(p_i,p_{2i+2})\rangle+\langle C(p_i,p_{2i+3})\rangle\cr
    &+\langle C(p_{2i+2},p_{2i+3})\rangle)\cr
    \leq&2\sum_{i=1}^{9\cdot(2^{k-1}-1)+3}n_i+|U_{\mFC}^k|.
\end{align}

\section{Non-contextual inequalities for some lattice hyper-graphs}

\begin{figure}
\includegraphics[scale=0.65]{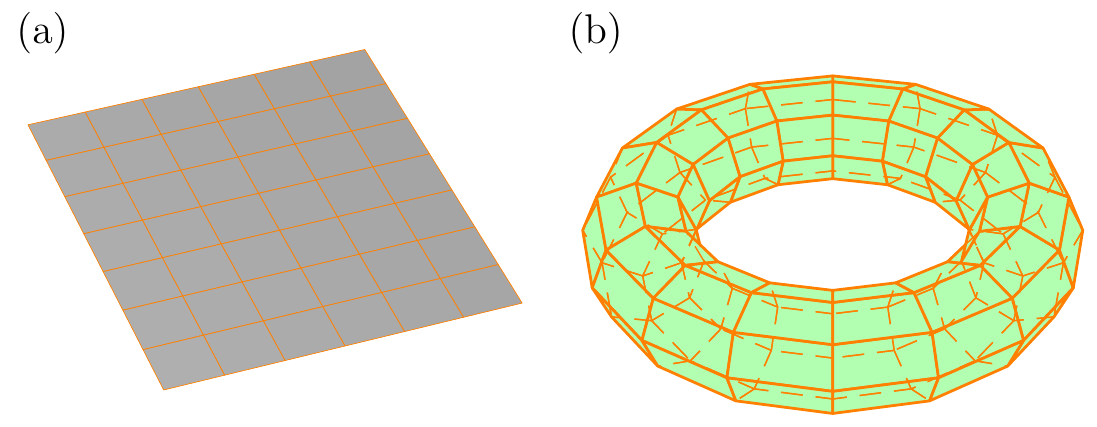} \caption{\label{period} (a)A two-dimensional square lattice  hyper-graph. (b) A two-dimensional torus lattice  hyper-graph.}
\end{figure}

In the end, let us consider two typical lattice hyper-graph families, see FIG.\ref{period}. 
We denote by $\mSL_{m_xm_y}$ and $\mTL_{m_xm_y}$ the square lattice  hyper-graph and the torus lattice  hyper-graph of $m_x\times m_y$ vertices respectively. We can get the classical upper bounds of their non-contextual inequalities by calculating the numbers of vertices, $|U_{\mSL}^{m_xm_y}|$ and $|U_{\mTL}^{m_xm_y}|$, in their maximal unconnected vertex sets $U_{\mSL}^{m_xm_y}$ and $U_{\mTL}^{m_xm_y}$. It is clear that $|U_{\mSL}^{m_xm_y}|$ and $|U_{\mTL}^{m_xm_y}|$ can be written as
\begin{align*}
    |U_{\mSL}^{m_xm_y}|=&\lceil\frac{{m_xm_y}}{2}\rceil;\cr
   |U_{\mTL}^{m_xm_y}|=&\lfloor\frac{\min\{m_x,m_y\}}{2}\rfloor\cdot\max\{m_x,m_y\}.
\end{align*}
Then, from Theorem 4, the non-contextual inequality for the square lattice  hyper-graph can be given by
\begin{align}\label{SLgraph}
    \langle \mSL_{m_xm_y}\rangle=&\sum_{i=1}^{m_x}\sum_{j=1}^{m_y}p_{i,j}
    +\sum_{i=1}^{m_x-1}\sum_{j=1}^{m_y}\langle C(p_{i,j},p_{i+1,j})\rangle\cr
    &+\sum_{i=1}^{m_x}\sum_{j=1}^{m_y-1}\langle C(p_{i,j},p_{i,j+1})\rangle\cr
    \leq&2(\sum_{i=1}^{m_x-1}\sum_{j=1}^{m_y}n_{i,j;i+1,j}+\sum_{i=1}^{m_x}\sum_{j=1}^{m_y-1}n_{i,j;i,j+1})\cr
    &+|U_{\mSL}^{m_xm_y}|,
\end{align}
where $p_{ij}$ is the vertex on the site $(i,j)$ and $n_{i,j;i+1,j}$ ($n_{i,j;i,j+1}$) is the  weight of the hyper-edge $(p_{i,j},p_{i+1,j})$ ($(p_{i,j},p_{i,j+1})$).

Likewise, for the torus lattice  hyper-graph,  as $m_x, m_y\geq3$, we can get the following non-contextual inequality
\begin{align}\label{TLgraph}
    \langle \mTL_{m_xm_y}\rangle=&\sum_{i=1}^{m_x}\sum_{j=1}^{m_y}p_{i,j}
    +\sum_{i=1}^{m_x}\sum_{j=1}^{m_y}(\langle C(p_{i,j},p_{i+1,j})\rangle\cr
    &+\langle C(p_{i,j},p_{i,j+1})\rangle)\cr
   \leq&2(\sum_{i=1}^{m_x}\sum_{j=1}^{m_y}n_{i,j;i+1,j}+\sum_{i=1}^{m_x}\sum_{j=1}^{m_y}n_{i,j;i,j+1})\cr
    &+|U_{\mTL}^{m_xm_y}|,
 \end{align}   
where $p_{m_x+1,j}\equiv p_{1,j}$ ($p_{i,m_y+1}\equiv  p_{i,1}$) and $n_{m_x,j;m_x+1,j}\equiv n_{m_x,j;1,j}$ ($n_{i,m_y;i,m_y+1}\equiv n_{i,m_y;i,1}$).

Other models such like cubic lattice hyper-graphs can also be discussed by using the same method. 

\section{Quantum violations}

To see the quantum violation for the non-contextual inequality for a $k$-vertex ordinary hyper-graph $G_k$, the key is to calculate the range of the eigenvalues for $\sum_{i=1}^kp_i$. As for any hyper-edge observable $C(p_i,p_j)$, from the view of the complete orthonormal bases,  the quantum expectation is strictly equal to $2n_{ij}$, where $n_{ij}$ is the corresponding hyper-edge weight. We denote by $\lambda_{\min}$ the minimal eigenvalue for $\sum_{i=1}^kp_i$. Then We have $\langle G_k\rangle_q^{\min}=2\sum_{i=1}^{|E|}n_i+\lambda_{\min}$, where the expression for $G_k$ can be found in Eq.(\ref{Rgraph}) and the notation $\langle\cdot \rangle_q$ represents the quantum expectation. If $\lambda_{\min}>|U|$,  then Eq.(\ref{Rgraph}) provide us a state-independent non-contextual inequality. An equivalent conclusion can also be found in Ref.\cite{Tang-yu}. For other cases,  it is at best a state-dependent non-contextual inequality.

\begin{figure}
\includegraphics[scale=0.7]{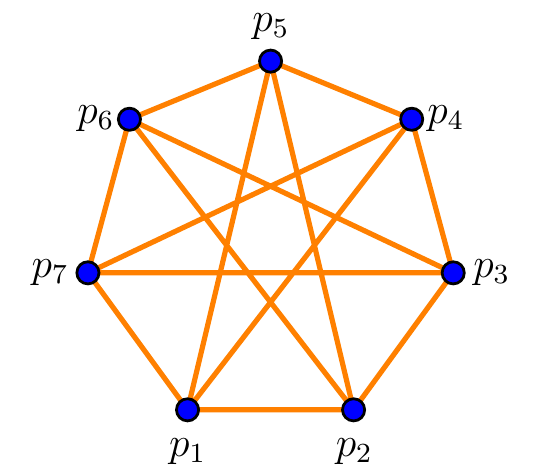} \caption{\label{7-vertex} A 7-vertex wheel hyper-graph $\mW_7$}.
\end{figure}

Besides calculating the eigenvalue for the sum of all the vertices in a hyper-graph, it seems that the relative size(compared with the vertex number of the hyper-graph) for a maximal unconnected vertex set of a hyper-graph may be one of the key factors in testing the quantum violation for the non-contextual inequality. Although sometimes it may not works very well, it can help us to get a preliminary estimation for the possibility of a quantum violation.
From this point of view, the non-contextual  inequality for the complete hyper-graph family in FIG.\ref{3-special-graph}-(a) seems to be the most possible case for a quantum violation. As the maximal unconnected vertex set is just a single vertex, and $\langle\sum_{i=1}^kp_i\rangle_q\geq\lambda_{\min}\geq1$ is easy to be satisfied. But the main shortcoming is that the number of the hyper-edges might be too large. To balance that, we try to choose the hyper-graphs with less hyper-edges, but might still have a state-independent quantum violation. Here we  give an example in FIG.\ref{7-vertex}

From Theorem 4, as $|U|=2$, the non-contextual inequality can be given by
\begin{align}\label{W7}
    \langle \mW_7\rangle&=\sum_{i=1}^7p_i+\sum_{i=1}^{7}(\langle C(p_i,p_{i+1})+\langle C(p_i,p_{i+3})\rangle)\cr
    &\leq2\sum_{i=1}^{14}n_i+2,
\end{align}
where $p_{8,9,10}\equiv p_{1,2,3}$ and $n_i$ is the weight for the $i$-th hyper-edge. To see the quantum violation, we choose $\{p_1,p_2,p_3,p_4\}$ to be the 4 core rays(vertices) of the Yu-Oh model\cite{yu-oh}, namely, $p_1,p_2,p_3,p_4$ are orienting to 4
vertices of a regular tetrahedron, and let $\{p_5,p_6,p_7\}$ be an approximate orthonormal basis (e.g. with an error of $\delta<0.01$), and also make sure that there are no parallel or antiparallel relationships for these rays,  then $\langle\sum_{i=1}^7p_i\rangle_q=\frac{4}{3}+1+O(\delta)\approx\frac{7}{3}>|U|$. And we can get a state-independent non-contextual inequality Eq.(\ref{W7}), with a reduction of $7$ hyper-edges compared with the extreme case of a 7-vertex complete hyper-graph.

We can see that the inequality approach based on hyper-graphs sometimes may help us to optimize the method for construction of a state-independent proof for quantum contextuality in Ref.\cite{Tang-yu} from the above example. In other words, by this method we may get a more economical proof for quantum contextuality with less auxiliary complete orthonormal bases (hyper-edges) but still in a state-independent manner.

Constraints for state-dependent non-contextual inequalities by other models referred in previous sections are listed in the following table,
\begin{center}
\begin{table}[htb]
\begin{tabular}{l|l}
  \hline\hline
  Models & Constraints for $\lambda_{\max}$  \\
   \hline
  $\mL_k$ & $\lambda_{\max}>\lceil\frac{k}{2}\rceil$ \\
  $\mmC_k$ & $\lambda_{\max}>\lfloor\frac{k}{2}\rfloor$ \\
  $\mFL_k$ & $\lambda_{\max}>\frac{4}{3}\cdot(2^k-2^{k\mod2-2})$ \\
  $\mFC_k$ & $\lambda_{\max}>2^k-1$  \\
  $\mSL_{m_xm_y}$ & $\lambda_{\max}>\lceil\frac{{m_xm_y}}{2}\rceil$  \\
  $\mTL_{m_xm_y}$ & $\lambda_{\max}>\lfloor\frac{\min\{m_x,m_y\}}{2}\rfloor\cdot\max\{m_x,m_y\}$  \\
  \hline\hline
\end{tabular}
\end{table}
\end{center}
where $\lambda_{\max}$ is the maximal eigenvalue for the corresponding $\sum_{i=1}^{|V|}p_i$ (or $\sum_{i=1}^{m_x}\sum_{j=1}^{m_y}p_{i,j}$) term.

\section{Conclusion and discussion}

We have discussed a general method for deriving the non-contextual inequalities based on the hyper-graphs for the qutrit systems.  Several interesting families of non-contextual inequalities are given. Our method can be applied to any hyper-graph by a subgraph decomposition relation. This relation might be very useful in looking for further interesting relations from other possible correlated structures. We also give the conditions for quantum violations of different types of non-contextual inequalities. Besides, our graphical methods might be helpful to improve the construction for state-independent proofs for quantum contextuality in our anther recent work\cite{Tang-yu}. Moreover, we notice that the mathematical structures of certain non-contextual inequalities  and the Hamiltonians for some systems in condensed matter physics are similar.  This might motivate us to give a further research on the link between them and try to look for a new method to learn some many-body physical systems.

\begin{acknowledgements}

This work is supported by the NNSF of China (Grant No. 11405120) and the Fundamental Research Funds for the Central Universities.
\end{acknowledgements}

\end{document}